\documentclass[letterpaper, 10 pt, onecolumn]{IEEEtran}  

\IEEEoverridecommandlockouts

\usepackage{amsmath} 
\usepackage{amssymb}  

\usepackage{amsthm}
\usepackage{graphicx}
\usepackage[space]{grffile}
\usepackage{cite}

\DeclareMathOperator*{\argmin}{arg\,min}

\expandafter\def\expandafter\normalsize\expandafter{%
    \normalsize
    \setlength\abovedisplayskip{4pt}
    \setlength\belowdisplayskip{4pt}
    \setlength\abovedisplayshortskip{4pt}
    \setlength\belowdisplayshortskip{4pt}
}

\title{\LARGE \bf
Convergence Analysis of Policy Iteration
}

\author{Ali Heydari$^1$
\thanks{$^{1}$Assistant Professor of Mechanical Engineering, South Dakota School of Mines and Technology, Rapid City, SD 57701, email: ali.heydari@sdsmt.edu.}}

\newtheorem{Thm}{Theorem} 
\newtheorem{Lem}{Lemma}

\newtheorem{Def}{Definition} 
\newtheorem{Not}{Notation} 
\newtheorem{Assumption}{Assumption} 

\begin{document}

\maketitle
\pagenumbering{arabic}
\pagestyle{plain}
\thispagestyle{plain}

\begin{abstract}
Adaptive optimal control of nonlinear dynamic systems with deterministic and known dynamics under a known undiscounted infinite-horizon cost function is investigated. Policy iteration scheme initiated using a stabilizing initial control is analyzed in solving the problem. The convergence of the iterations and the optimality of the limit functions, which follows from the established uniqueness of the solution to the Bellman equation, are the main results of this study. Furthermore, a theoretical comparison between the speed of convergence of policy iteration versus value iteration is presented. Finally, the convergence results are extended to the case of multi-step look-ahead policy iteration.
\end{abstract}

\section{Introduction}
This short study investigates the convergence of the policy iteration (PI) as one of the schemes in implementation of adaptive/approximate dynamic programming (ADP), sometimes referred to by reinforcement learning (RL) or neuro-dynamic programming (NDP), \cite{Watkins}-\cite{Heydari_TCYB}. 

Compared to its alternative, i.e., value iteration (VI), 
the PI calls for a higher computational load per iteration, due to a `full backup' as opposed to a `partial backup' in VI, \cite{LewisContSystMag}. 
However, the PI has the advantage that the control under evolution remains stabilizing \cite{Liu_PI}, hence, it is more suitable for online implementation, i.e., adapting the control `on the fly'. 
The convergence analyses for PI with continuous state and control spaces and an undiscounted cost function are given in \cite{Liu_PI}. The results presented in this study however, are from a different viewpoint with different assumptions and lines of proofs.
Moreover, interested readers are referred to the results from a simultaneous research (at least in terms of the availability of the results to the public) presented in \cite{Bertsekas_VI_PI_2015}, which are the closest to the first two theorems of this study. 

This study establishes the convergence of the PI to the solution to the optimal control problem with known deterministic dynamics. Moreover, given the faster convergence of PI compared with VI which can be observed in numerical implementations, some theoretical results are presented which compare the rates of convergences.
Finally, the multi-step look-ahead variation of PI, \cite{Bertsekas2012}, is analyzed and its convergence is established.

\section{Problem Formulation} \label{ProblemFormulation}
The discrete-time nonlinear system given by
\begin{equation}
x_{k+1}=f(x_k,u_k), k \in \mathbb{N}, \label{Dynamics}
\end{equation}
is subject to control, where (possibly discontinuous) function $f:\mathbb{R}^n \times \mathbb{R}^m \to \mathbb{R}^n$ is known, the state and control vectors are denoted with $x$ and $u$, respectively, and $f(0,0)=0$. Positive integers $n$ and $m$ denote the dimensions of the continuous state space $\mathbb{R}^n$ and the (possibly discontinuous) control space $\mathcal{U} \subset \mathbb{R}^m$, respectively, sub-index $k$ represents the discrete time index, and the set of non-negative integer numbers is denoted with $\mathbb{N}$. 
The cost function subject to minimization is given by 
\begin{equation}
J=\sum_{k=0}^\infty {U(x_k,u_k)}, \label{CostFunction}
\end{equation}
where the \textit{utility function} $U:\mathbb{R}^n \times \mathcal{U} \to \mathbb{R}_+$ is positive semi-definite with respect to the first input, and positive definite with respect to its second input. The set of non-negative real numbers is denoted with $\mathbb{R}_+$.

Selecting an initial feedback \textit{control policy} $h:\mathbb{R}^n \to \mathcal{U}$, i.e., $u_k = h(x_k)$, the adaptive optimal control problem is \textit{updating/adapting} the control policy such that cost function (\ref{CostFunction}) is minimized. The minimizing control policy is called the \textit{optimal control policy} and denoted with $h^*(.)$.

\begin{Not} The state trajectory initiated from the initial state $x_0$ and propagated using the control policy $h(.)$ is denoted with $x_k^h, \forall k \in \mathbb{N}$. In other words, $x_0^h:=x_0$ and $x_{k+1}^h = f\big(x_k^h, h(x_k^h)\big), \forall k \in \mathbb{N}$.
\end{Not} 

\begin{Def} \label{AsymStability_Definition}
The control policy $h(.)$ is defined to be {asymptotically stabilizing} within a domain if $lim_{k\to\infty} x_k^h = 0$, for every initial $x_0^h$ selected within the domain, \cite{Khalil}.   
\end{Def} 

\begin{Def} \label{Def1}
The set of admissible control policies (within a compact set), denoted with $\mathcal{H}$, is defined as the set of policies $h(.)$ that asymptotically stabilize the system within the set and 
their respective \textit{`cost-to-go'} or \textit{`value function'}, denoted with $V_h:\mathbb{R}^n \to \mathbb{R}_+$ and defined by 
\begin{equation}
V_h(x_0):=\sum_{{k}=0}^\infty {U\big(x_{k}^h, h(x_{k}^h)\big)}, \label{ValueFunction_of_h}
\end{equation}
is upper bounded within the compact set by a continuous function $\bar{V}:\mathbb{R}^n \to \mathbb{R}_+$ where $\bar{V}(0)=0$.
\end{Def}

If the value function itself is continuous, the upper boundedness by $\bar{V}(.)$ is trivially met, through selecting $\bar{V}(.) = V_h(.)$. Note that the continuity of the upper bound in the compact set leads to its finiteness within the set, and hence, the finiteness of the value function. This is a critical feature for the value function and hence, the control policy.

\begin{Assumption} \label{Assum_ExistingAdmissibleCont}
There exists at least one admissible control policy for the given system within a connected and compact set $\Omega \subset \mathbb{R}^n$ containing the origin.
\end{Assumption}

\begin{Assumption} \label{Assum_InvariantSet}
The intersection of the set of n-vectors $x$ at which $U(x,0) = 0$ with the invariant set of $f(.,0)$ only contains the origin. 
\end{Assumption}

Assumption \ref{Assum_ExistingAdmissibleCont} leads to the conclusion that the value function associated with the \textit{optimal} control policy is finite at any point in $\Omega$, as it will not be greater than $V_h(.)$ at that point, for any admissible control policy $h(.)$. 
Assumption \ref{Assum_InvariantSet} implies that the optimal control policy will be asymptotically stabilizing, as there is no non-zero state trajectory that can `hide' somewhere without convergence to the origin. Given these two assumptions, it is concluded that the optimal control policy is an admissible policy, i.e., $h^*(.) \in \mathcal{H}$.

\section{ADP-based Solutions}
The Bellman equation \cite{Bertsekas_NDP}, given below, provides the \textit{optimal value function}
\begin{equation}
		V^*(x) = \min_{u\in\mathcal{U}} \Big( U\big(x,u\big) + V^*\big(f\big(x,u\big)\big)\Big), \label{Bellman_eq1}
\end{equation}
which once obtained, leads to the solution to the problem, through
\begin{equation}
		h^*(x) = \argmin_{u\in\mathcal{U}} \Big( U\big(x,u\big) + V^*\big(f\big(x,u\big)\big)\Big). \label{Bellman_eq2}
\end{equation}
But, this is mathematically impracticable for general nonlinear systems, \cite{Bertsekas_NDP}.
Policy iteration (PI) provides a learning algorithms for training a function approximator or forming a lookup table, for approximating the solution \cite{Howard_PI_MDP, Bertsekas2012, Werbos2012}. This approximation is done within a compact and connected set, containing the origin, called the domain of interest and denoted with $\Omega$.

Starting with an initial admissible control policy, denoted with $h^0(.)$, one iterates through the \textit{policy evaluation equation} given by
\begin{equation}
		V^i(x) = U\big(x,h^i(x)\big) + V^i\Big(f\big(x,h^i(x)\big)\Big), \forall x \in \Omega, \label{PI_PolicyEval}
\end{equation}
and the \textit{policy update equation} given by
\begin{equation}
		h^{i+1}(x) = \argmin_{u\in\mathcal{U}} \Big( U\big(x,u\big) + V^i\big(f\big(x,u\big)\big)\Big), \forall x \in \Omega, \label{PI_PolicyUpdate}
\end{equation}
for $i=0,1,...$ until they converge, in PI. Each of these equations may be evaluated at different points in $\Omega$, for obtaining the \emph{targets} for training the respective function approximators.

\section{Convergence Analysis of Policy Iteration}

Given the fact that Eqs. (\ref{PI_PolicyEval}) and (\ref{PI_PolicyUpdate}) are iterative equations, the following questions arise. 1- Does the iterations converge? 2- If they converge, are the limit functions optimal? This section is aimed at answering these two questions. 
Initially the following two lemmas are presented. 

\begin{Lem} \label{Lemma_1} Given admissible control policies $h(.)$ and $g(.)$, if 
\begin{equation}
U\big(x,h(x)\big) + V_g\Big(f\big(x,h(x)\big)\Big) \leq V_g(x), \forall x \in \Omega, \label{eq_Lemma1_eq1}
\end{equation}
then $V_h(x) \leq V_g(x), \forall x \in \Omega$.
\end{Lem}
\textit{Proof}: Evaluating (\ref{eq_Lemma1_eq1}) at $x_0^h \in \Omega,$ one has
\begin{equation}
U\big(x_0^h,h(x_0^h)\big) + V_g\big(x_1^h\big) \leq V_g(x_0^h), \forall x_0^h \in \Omega. \label{eq_Lemma1_eq2}
\end{equation}
Also, evaluating (\ref{eq_Lemma1_eq1}) at $x_1^h$ leads to
\begin{equation}
U\big(x_1^h,h(x_1^h)\big) + V_g\big(x_2^h\big) \leq V_g(x_1^h), \forall x_1^h \in \Omega. \label{eq_Lemma1_eq3}
\end{equation}
Using (\ref{eq_Lemma1_eq3}) in (\ref{eq_Lemma1_eq2}) leads to
\begin{equation}
U\big(x_0^h,h(x_0^h)\big) + U\big(x_1^h,h(x_1^h)\big) + V_g\big(x_2^h\big) \leq V_g(x_0^h), \forall x_0^h \in \Omega. \label{eq_Lemma1_eq4}
\end{equation}
Repeating this process for $N-2$ more times leads to 
\begin{equation}
\sum_{k=0}^{N-1} U\big(x_k^h,h(x_k^h)\big) + V_g\big(x_{N}^h\big) \leq V_g(x_0^h), \forall x_0^h \in \Omega. \label{eq_Lemma1_eq5}
\end{equation}
Letting $N \to \infty$ and given $V_g(x) \geq 0, \forall x$, which hence can be dropped from the left hand side, inequality (\ref{eq_Lemma1_eq5}) leads to $V_h(x) \leq V_g(x), \forall x \in \Omega$, by definition of $V_h(.)$.
\qed

\vspace{10pt}

Assuming two admissible control policies $h(.)$ and $g(.)$, Lemma \ref{Lemma_1} simply shows that if applying $h(.)$ at the first time step and applying $g(.)$ for infinite number of times in the future, leads to a cost not greater than only applying $g(.)$, then, the value function of $h(.)$ also will not be greater than that of $g(.)$, at any point.

\begin{Lem} \label{Lemma_2} Given admissible control policies $h(.)$ and $g(.)$, if 
\begin{equation}
V_h(x) < V_g(x), \exists x \in \Omega, \label{eq_Lemma2_eq1}
\end{equation}
then
\begin{equation}
U\big(x,h(x)\big) + V_g\Big(f\big(x,h(x)\big)\Big) < V_g(x), \exists x \in \Omega. \label{eq_Lemma2_eq2}
\end{equation}
\end{Lem}
\textit{Proof}: The proof is done by contradiction. Assume that (\ref{eq_Lemma2_eq2}) does not held, i.e.,
\begin{equation}
U\big(x_0^h,h(x_0^h)\big) + V_g\big(x_1^h\big) \geq V_g(x_0^h), \forall x_0^h \in \Omega, \label{eq_Lemma2_eq3}
\end{equation}
which leads to 
\begin{equation}
U\big(x_1^h,h(x_1^h)\big) + V_g\big(x_2^h\big) \geq V_g(x_1^h), \forall x_1^h \in \Omega. \label{eq_Lemma2_eq4}
\end{equation}
Using (\ref{eq_Lemma2_eq4}) in (\ref{eq_Lemma2_eq3}) leads to
\begin{equation}
U\big(x_0^h,h(x_0^h)\big) + U\big(x_1^h,h(x_1^h)\big) + V_g\big(x_2^h\big) \geq V_g(x_0^h), \forall x_0^h \in \Omega. \label{eq_Lemma2_eq5}
\end{equation}
Repeating this process for $N-2$ more times, one has
\begin{equation}
\sum_{k=0}^{N-1} U\big(x_k^h,h(x_k^h)\big) + V_g\big(x_{N}^h\big) \geq V_g(x_0^h), \forall x_0^h \in \Omega. \label{eq_Lemma2_eq6}
\end{equation}
Let $N \to \infty$. Given the admissibility of $g(.)$ and $h(.)$, one has $V_g(x_{N}^h) \to 0, \forall x_0^h,$ as $N\to\infty$. The reason is $\lim_{N\to\infty} x_N^h\to 0$ and the continuity of the upper bound of $V_g(.)$, per the admissibility of $g(.)$.
Therefore, Inequality (\ref{eq_Lemma2_eq6}) contradicts (\ref{eq_Lemma2_eq1}), because, the second term in the left hand side of (\ref{eq_Lemma2_eq6}) can be made arbitrarily small. Hence, (\ref{eq_Lemma2_eq6}) leads to $V_h(x) \geq V_g(x), \forall x \in \Omega$ \footnote{This conclusion can also be made using another contradiction argument, through (\ref{eq_Lemma2_eq1}) which leads to $V_h(x_0) + \epsilon = V_g(x_0), \exists x_0 \in \Omega$, for some $\epsilon=\epsilon(x_0) > 0$. Then, selecting large enough $N$ such that $V_g(x_{N}^h) < \epsilon$, inequality (\ref{eq_Lemma2_eq6}) contradicts (\ref{eq_Lemma2_eq1}).}, which contradicts (\ref{eq_Lemma2_eq1}), hence, (\ref{eq_Lemma2_eq3}) cannot hold. This completes the proof.
\qed

\vspace{10pt}

In simple words, Lemma \ref{Lemma_2} shows that if the value function of $h(.)$ is less than that of $g(.)$ at least at one $x$, then, the cost of applying $h(.)$ only at the first step and applying $g(.)$ for the rest of the steps also will be less than the cost of only applying $g(.)$ throughout the horizon, at least at one $x$.
This result leads to the uniqueness of the solution to the Bellman equation (\ref{Bellman_eq1}), as shown in the next theorem.

\begin{Thm} \label{Thm_Uniq} The Bellman equation given by (\ref{Bellman_eq1}) has a unique solution in $\Omega$.
\end{Thm}
\textit{Proof}: The proof is by contradiction. Assume that there exists some $V_h(.)$ that satisfies
\begin{equation}
		V_h(x) = \min_{u\in\mathcal{U}} \Big( U\big(x,u\big) + V_h\big(f\big(x,u\big)\big)\Big), \forall x\in\Omega, \label{eq_Thm_Uniq_eq1}
\end{equation}
while, $V^*(x) < V_h(x), \exists x \in\Omega$, in other words $h^*(x) \neq h(x), \exists x \in\Omega$, where 
\begin{equation}
h(x) := \argmin_{u\in\mathcal{U}} \Big( U\big(x,u\big) + V_h\big(f\big(x,u\big)\big)\Big), \forall x\in\Omega. \label{eq_Thm_Uniq_eq2}
\end{equation}
Using Lemma \ref{Lemma_2}, inequality $V^*(x) < V_h(x), \exists x \in\Omega,$ leads to
\begin{equation}
U\big(x,h^*(x)\big) + V_h\Big(f\big(x,h^*(x)\big)\Big) < V_h(x) = U\big(x,h(x)\big) + V_h\Big(f\big(x,h(x)\big)\Big), \exists x \in \Omega. \label{eq_Thm_Uniq_eq3}
\end{equation}
But, (\ref{eq_Thm_Uniq_eq3}) contradicts (\ref{eq_Thm_Uniq_eq2}). Hence, $h^*(x) = h(x), \forall x \in\Omega$, and therefore, $V^*(x) = V_h(x), \forall x \in\Omega$. 
\qed.

\vspace{10pt}
The next step is the proof of convergence of PI.

\begin{Thm} \label{Thm_Conv_PI} The policy iteration given by equations (\ref{PI_PolicyEval}) and (\ref{PI_PolicyUpdate}) converges monotonically to the optimal solution in $\Omega$.
\end{Thm}
\textit{Proof}: The first step is showing the monotonicity of the sequence of value functions $\{V^i(x)\}_{i=0}^{\infty}$ generated using the PI equations. By (\ref{PI_PolicyUpdate}), one has
\begin{equation}
		U\big(x,h^{i+1}(x)\big) + V^i\Big(f\big(x,h^{i+1}(x)\big)\Big) \leq V^i(x), \forall x \in \Omega. \label{eq_Thm_Conv_PI_eq1}
\end{equation}
Using Lemma \ref{Lemma_2}, the former inequality leads to
\begin{equation}
		V^{i+1}(x) \leq V^i(x), \forall x \in \Omega, \label{eq_Thm_Conv_PI_eq2}
\end{equation}
for any selected $i$. Hence, $\{V^i(x)\}_{i=0}^{\infty}$ is pointwise decreasing. On the other hand, it is lower bounded by the optimal value function. Hence, it converges, \cite{Rudin}. Denoting the limit value function and the limit control policy with $V^{\infty}(.)$ and $h^{\infty}(.)$, respectively, they satisfy PI equations
\begin{equation}
		V^{\infty}(x) = U\big(x,h^{\infty}(x)\big) + V^{\infty}\big(f\big(x,h^{\infty}(x)\big)\big), \forall x\in\Omega, \label{eq_Thm_Conv_PI_eq2}
\end{equation}
and
\begin{equation}
		h^{\infty}(x) = \argmin_{u\in\mathcal{U}} \Big( U\big(x,u\big) + V^{\infty}\big(f\big(x,u\big)\big)\Big), \forall x\in\Omega, \label{eq_Thm_Conv_PI_eq3}
\end{equation}
hence,
\begin{equation}
		V^{\infty}(x) = \min_{u\in\mathcal{U}} \Big( U\big(x,u\big) + V^{\infty}\big(f\big(x,u\big)\big)\Big), \forall x\in\Omega. \label{eq_Thm_Conv_PI_eq4}
\end{equation}
Eq. (\ref{eq_Thm_Conv_PI_eq4}) is the Bellman equation, which per Theorem \ref{Thm_Uniq} has a unique solution. Hence, $V^{\infty}(.)=V^*(.)$ everywhere in $\Omega$. This completes the proof.
\qed
\vspace{10pt}

\begin{Thm} \label{Thm_Adm_PI} The control policies at the iterations of the policy iteration given by equations (\ref{PI_PolicyEval}) and (\ref{PI_PolicyUpdate}) remain admissible in $\Omega$.
\end{Thm}
\textit{Proof}: Given the requirements for admissibility, one needs to show that each policy is asymptotically stabilizing and its respective value function is upper bounded by a continuous function which passes through the origin. The latter follows from the monotonicity of the sequence of value functions under VI, established in Theorem \ref{Thm_Conv_PI}, since $h^0(.)$ is admissible. The former, also follows from this monotonicity, as no state trajectory can hide in the set at which the utility function is zero, without convergence to the origin, per Assumption \ref{Assum_InvariantSet}. In other words, in order for its value function to be bounded, the policy needs to steer the trajectory towards the origin.
\qed
\vspace{10pt}

\section{Comparison between Policy and Value iterations}
Value iteration (VI), as an alternative to PI, is conducted using an initial guess $W^0(.)$ and iterating through the \textit{policy update equation} given by
\begin{equation}
		g^{i}(x) = \argmin_{u\in\mathcal{U}} \Big( U\big(x,u\big) + W^i\big(f\big(x,u\big)\big)\Big), \forall x \in \Omega, \label{VI_PolicyUpdate}
\end{equation}
and the \textit{value update equation} 
\begin{equation}
		W^{i+1}(x) = U\big(x,g^i(x)\big) + W^i\Big(f\big(x,g^i(x)\big)\Big), \forall x \in \Omega. \label{VI_ValueUpdate2}
\end{equation}
The two former equations can be merged into
\begin{equation}
		W^{i+1}(x) = \min_{u\in\mathcal{U}} \Big( U\big(x,u\big) + W^i\big(f\big(x,u\big)\big)\Big), \forall x \in \Omega. \label{VI_ValueUpdate}
\end{equation}
for $i=0,1,...$, where notations $W^i(.)$ and $g^i(.)$ are used for the value function and the control policy resulting from the VI, respectively, for clarity. The convergence proof of VI is not the subject of this study and can be found in many references including \cite{Lincol_RelaxingDynProg}, \cite{AlTamimi}, and \cite{Heydari_TCYB}. 

The VI has the advantage of not requiring an admissible control as the initial guess. The PI, however, has the advantage that the control policies subject to evolution remain stabilizing for the system. It was shown in \cite{Heydari_SVI} that if the VI is also initiated using an admissible initial guess, the control policies remain stabilizing. Therefore, starting with an admissible guess, the VI and PI seem to be similar in terms of stability. The computational load per iteration in VI is significantly less than that of PI, due to needing to do a simple recursion in VI using (\ref{VI_ValueUpdate2}), called a `partial backup' in \cite{LewisContSystMag}, as compared with solving an equation in PI, namely, Eq. (\ref{PI_PolicyEval}), which is a `full backup', \cite{LewisContSystMag}. However, in practice, it can be seen that the PI converges much faster than the VI, in terms of the number of iterations. This section is aimed at providing some analytical results confirming this observation. 

\begin{Thm} \label{Thm_PI_vs_VI} If $V^0(.)=W^0(.)$ is calculated using as admissible control policy, the policy iteration given by equations (\ref{PI_PolicyEval}) and (\ref{PI_PolicyUpdate}) converges not slower than the value iteration given by equations (\ref{VI_PolicyUpdate}) and (\ref{VI_ValueUpdate2}), in $\Omega$.
\end{Thm}
\textit{Proof}: Given the convergence of both schemes to the unique $V^*(.)$, the claim is proved by showing that $V^i(x) \leq W^i(x), \forall x\in\Omega, \forall i \in \mathbb{N}$. From $V^0(.)=W^0(.)$ one has $h^1(.)=g^0(.)$. Hence,
\begin{equation}
\begin{split}
		W^{1}(x) = U\big(x,g^0(x)\big) + W^0\Big(f\big(x,g^0(x)\big)\Big) & = U\big(x,h^1(x)\big) + V^0\Big(f\big(x,h^1(x)\big)\Big) \geq \\
		&U\big(x,h^1(x)\big) + V^1\Big(f\big(x,h^1(x)\big)\Big) = V^1(x), \forall x \in \Omega, \label{eq_Thm_PI_vs_VI_eq1}
\end{split}
\end{equation}
where the inequality is due to the monotonically decreasing nature of $\{V^i(x)\}_{i=0}^{\infty}$ established in Theorem \ref{Thm_Conv_PI}. Therefore, $W^{1}(x) \geq V^{1}(x), \forall x$. Now, assume that $W^{i}(x) \geq V^{i}(x), \forall x$, for some $i$. Then, 
\begin{equation}
\begin{split}
		W&^{i+1}(x)= U\big(x,g^i(x)\big) + W^i\Big(f\big(x,g^i(x)\big)\Big) \geq U\big(x,g^i(x)\big) + V^i\Big(f\big(x,g^i(x)\big)\Big) \geq \\
		&U\big(x,h^{i+1}(x)\big) + V^i\Big(f\big(x,h^{i+1}(x)\big)\Big) \geq U\big(x,h^{i+1}(x)\big) + V^{i+1}\Big(f\big(x,h^{i+1}(x)\big)\Big) = V^{i+1}(x), \forall x \in \Omega, \label{eq_Thm_PI_vs_VI_eq2}
\end{split}
\end{equation}
The first inequality is due to the assumed $W^{i}(x) \geq V^{i}(x), \forall x$. The second inequality is due to the fact that $h^{i+1}(.)$ is the minimizer of the term subject to comparison, and the last inequality is due to the monotonicity of $\{V^i(x)\}_{i=0}^{\infty}$. Hence, $W^{i+1}(x) \geq V^{i+1}(x), \forall x\in\Omega,$ and the claim is proved by induction.
\qed
\vspace{10pt}

It should be noted that the result given in the former theorem is probably very conservative, as it only shows that the convergence of the PI will not be `slower' than that of the VI.

\section{Convergence Analysis of Multi-step Look-ahead Policy Iteration}
Multi-step Look-ahead Policy Iteration (MLPI), \cite{Bertsekas_NDP}, is a variation of PI, given by the policy evaluation equation (\ref{PI_PolicyEval}) repeated below
$$
		V^i(x_0) = U\big(x_0,h^i(x_0)\big) + V^i\Big(f\big(x_0,h^i(x_0)\big)\Big), \forall x_0 \in \Omega,
$$
and the new \textit{policy update equation} with $n$-step look-ahead ($n \in \mathbb{N}, n > 0$) given by
\begin{equation}
		h^{i+1}(x_0^h) = \argmin_{h\in\mathcal{H}} \Big( \sum_{k=0}^{n-1} U\big(x_k^h,h(x_k^h)\big) + V^i\big(x_n^h\big)\Big), \forall x_0^h=x_0 \in \Omega. \label{MLPI_PolicyUpdate}
\end{equation}
It can be seem that the regular PI is a special case of the MLPI with $n=1$, \cite{Bertsekas_NDP}. It is not surprising to expect the MLPI to converge faster than the regular PI, as in the extreme case that $n \to \infty$, the optimal solution will be calculated in one iteration, using (\ref{MLPI_PolicyUpdate}), i.e., the iterations converge to the optimal solution after the very first iteration. The rest of this section provides the convergence analysis for MLPI, for $1 < n < \infty$.

\begin{Thm} \label{Thm_Conv_MLPI} The multi-step look-ahead policy iteration given by equations (\ref{PI_PolicyEval}) and (\ref{MLPI_PolicyUpdate}) converges monotonically to the optimal solution in $\Omega$.
\end{Thm}
\textit{Proof}: The proof is similiar to the proof of Theorem \ref{Thm_Conv_PI}. Initially it is shown that the sequence of value functions $\{V^i(x)\}_{i=0}^{\infty}$ generated using the MLPI is monotonically decreasing. By (\ref{MLPI_PolicyUpdate}), one has
\begin{equation}
		\sum_{k=0}^{n-1} U\big(x_k^{h^{i+1}},h^{i+1}(x_k^{h^{i+1}})\big) + V^i\big(x_n^{h^{i+1}}\big) \leq V^i(x_0^{h^{i+1}}), \forall x_0^{h^{i+1}} = x_0 \in \Omega, \label{eq_Thm_Conv_MLPI_eq1}
\end{equation}
which is the consequence of $h^{i+1}(.)$ being the minimizer of the left hand side of the former inequality. Using the line of proof in the proof of Lemma \ref{Lemma_2}, inequality (\ref{eq_Thm_Conv_MLPI_eq1}) may be repeated in itself for infinite number of times to get 
\begin{equation}
		V^{i+1}(x) \leq V^i(x), \forall x \in \Omega, \label{eq_Thm_Conv_MLPI_eq2}
\end{equation}
which is valid for any selected $i$. Hence, $\{V^i(x)\}_{i=0}^{\infty}$ under the MLPI is pointwise decreasing. It is also lower bounded by the optimal value function, therefore, converges, \cite{Rudin}. Denoting the limit value function and the limit control policy with $V^{\infty}(.)$ and $h^{\infty}(.)$, respectively, they satisfy the MLPI equations
\begin{equation}
		V^{\infty}(x_0) = U\big(x_0,h^{\infty}(x_0)\big) + V^{\infty}\big(f\big(x_0,h^{\infty}(x_0)\big)\big), \forall x_0\in\Omega, \label{eq_Thm_Conv_MLPI_eq2}
\end{equation}
and
\begin{equation}
h^{\infty}(x_0^h) = \argmin_{h\in\mathcal{H}} \Big( \sum_{k=0}^{n-1} U\big(x_k^h,h(x_k^h)\big) + V^{\infty}\big(x_n^h\big)\Big), \forall x_0^h=x_0 \in \Omega, \label{eq_Thm_Conv_MLPI_eq3}
\end{equation}
hence,
\begin{equation}
V^{\infty}(x_0) = \min_{h\in\mathcal{H}} \Big( \sum_{k=0}^{n-1} U\big(x_k^h,h(x_k^h)\big) + V^{\infty}\big(x_n^h\big)\Big), \forall x_0^h=x_0 \in \Omega. \label{eq_Thm_Conv_MLPI_eq4}
\end{equation}
Eq. (\ref{eq_Thm_Conv_MLPI_eq4}) is the $n$-step look-ahead version of the Bellman equation (\ref{Bellman_eq1}) and $V^*(.)$ satisfies it, by definition. It can be proved that this equation also has a unique solution, which is $V^*(.)$, using the line of proof in Lemma \ref{Lemma_2} and Theorem \ref{Thm_Uniq}. To this end, assume that 
\begin{equation}
V^*(x) < V^{\infty}(x), \exists x \in \Omega, \label{eq_Thm_Conv_MLPI_eq5}
\end{equation}
hence, $h^*(x) \neq h^{\infty}(x), \exists x \in \Omega$. Inequality (\ref{eq_Thm_Conv_MLPI_eq5}) leads to
\begin{equation}
\sum_{k=0}^{n-1} U\big(x_k^{h^*},h^*(x_k^{h^*})\big) + V^{\infty}\big(x_n^{h^*}\big) < V^{\infty}(x_0^{h^*}), \exists x_0^{h^*} \in \Omega. \label{eq_Thm_Conv_MLPI_eq6}
\end{equation}
Otherwise, one has 
\begin{equation}
\sum_{k=0}^{n-1} U\big(x_k^{h^*},h^*(x_k^{h^*})\big) + V^{\infty}\big(x_n^{h^*}\big) \geq V^{\infty}(x_0^{h^*}), \forall x_0^{h^*} \in \Omega, \label{eq_Thm_Conv_MLPI_eq7}
\end{equation}
which repeating it in itself for unlimited number of times, and considering the fact that $V^{\infty}(.)$ in the left hand side can be made arbitrarily small after a large enough number of repetitions, leads to 
\begin{equation}
V^*\big(x_0^{h^*}\big) \geq V^{\infty}(x_0^{h^*}), \forall x_0^{h^*} \in \Omega. \label{eq_Thm_Conv_MLPI_eq8}
\end{equation}
But, (\ref{eq_Thm_Conv_MLPI_eq8}) contradicts (\ref{eq_Thm_Conv_MLPI_eq5}), hence, (\ref{eq_Thm_Conv_MLPI_eq5}) leads to (\ref{eq_Thm_Conv_MLPI_eq6}). But, (\ref{eq_Thm_Conv_MLPI_eq6}) contradicts $h^*(x) \neq h^{\infty}(x), \exists x \in \Omega$, per (\ref{eq_Thm_Conv_MLPI_eq3}), given $h^*(.) \in \mathcal{H}$. Therefore, (\ref{eq_Thm_Conv_MLPI_eq5}) cannot hold and $V^*(x) = V^{\infty}(x), \forall x \in \Omega$, which completes the proof.
\qed

\vspace{10pt}

\section{Conclusions}
The convergence of the policy iteration scheme to the solution of optimal control problems was analyzed. The speed of convergence of the policy iteration was shown to be not slower than that of the value iteration. Finally, the convergence of the multi-step look-ahead policy iteration to the optimal solution was established.

\bibliographystyle{ieeetr}

\end{document}